\documentclass[a4paper,fleqn,usenatbib]{mnras}

\usepackage{ae,aecompl}

\usepackage{graphicx}	
\usepackage{amsmath}	
\usepackage{amssymb}	

\newcommand{\music}{\textsc{music}}
\newcommand{\ramses}{\textsc{ramses}}

\usepackage{txfonts}
\usepackage[T1]{fontenc}

\usepackage{microtype}

\def\rv{r_{\rm 500}}

\title[Structure of the most massive galaxy clusters]{Internal dark matter structure of the most massive galaxy clusters}

\author[A. M. C. Le Brun et al.]{
A.~M.~C.~Le~Brun$^{1,2}$\thanks{E-mail: amandine.le-brun@cea.fr (AMCLB)},
M. Arnaud$^{1,2},$
G. W. Pratt$^{1,2}$
and R. Teyssier$^{3}$ 
\\
$^{1}$IRFU, CEA, Université Paris-Saclay, F-91191 Gif-sur-Yvette, France
\\
$^{2}$Université Paris Diderot, AIM, Sorbonne Paris Cité, CEA, CNRS, F-91191 Gif-sur-Yvette, France
\\
$^{3}$ Institute for Computational Science, University of Z\"urich, CH-8057 Z\"urich, Switzerland 
}

\date{Accepted 2017 September 21. Received 2017 August 25; in original form 2017 July 10.}

\pubyear{2017}

\begin{document}
\label{firstpage}
\pagerange{\pageref{firstpage}--\pageref{lastpage}}
\maketitle

\begin{abstract}
We investigate the evolution of the dark matter density profiles of the most massive galaxy clusters in the Universe. Using a `zoom-in' procedure on a large suite of cosmological simulations of total comoving  volume of  $3\,(h^{-1}\,\rm Gpc)^3$,  we study the 25 most massive clusters in  four redshift slices from $z\sim 1$ to the present. The minimum mass  is $M_{500} > 5.5 \times 10^{14}$ M$_{\odot}$  at $z=1$. Each system has more than two million particles within $\rv$. Once scaled to the critical density at each redshift, the dark matter profiles within $\rv$ are strikingly similar from $z\sim1$ to the present day, exhibiting a low dispersion of 0.15 dex, and showing little evolution with redshift in the radial logarithmic slope and scatter. They have the running power law shape typical of the NFW-type profiles, and their inner structure,  resolved to $3.8\,h^{-1}$ comoving kpc at $z=1$, shows no signs of converging to an asymptotic slope. Our results suggest that this type of  profile is already in place at $z>1$ in the highest-mass haloes in the Universe, and that it remains exceptionally robust to merging activity. 
\end{abstract}

\begin{keywords}
galaxies : clusters: general -- galaxies: evolution -- galaxies : structure -- cosmology: large-scale structure of the Universe -- cosmology: miscellaneous 
\end{keywords}




\section{Introduction}

While the theoretical paradigm for the formation and evolution of large-scale structure in a dark-matter-dominated universe was described over 35 years ago \citep[e.g.][]{Peebles1980}, it is only recently that observations have definitively established the conceptual framework within which this process takes place  \citep[e.g.][]{WMAP9,Planck2013,Planck2015}. Today, the hierarchical collapse of dark matter into clumps or `haloes' in the dark-energy dominated cold dark matter ($\Lambda$CDM) model is a cornerstone of our understanding of the formation of galaxies, galaxy groups and clusters. 

The most massive galaxy clusters, defined here as those with $M_{500} > 5 \times 10^{14}$ M$_{\odot}$, occupy a unique position in the cosmic hierarchy\footnote{see below for definition of $M_{500}$}. They are dark matter-dominated, except in the very centre. Their deep potential wells ensure that gravity is the dominant mechanism driving their evolution, leaving their observed properties least affected by the complicated non-gravitational processes linked to galaxy formation \citep[e.g.][]{Cui2014,Martizzi2014,Velliscig2014}. They are observable up to high redshift, and complementary techniques can be used to probe their internal structure and measure their scaling properties. For these reasons, they are ideal objects with which to test theories of structure formation. 

The advent of Sunyaev-Zel'dovich (SZ) surveys has led to an explosion in  the number of known high-redshift clusters \citep{PSZ1,has13,Bleem2015,PSZ2}. Many of these are high mass, and recent results have demonstrated the feasibility of obtaining high-quality structural and scaling information from these objects \citep[e.g.][]{Bartalucci2017,sch16}.  In particular, X-ray observations of these bright systems can probe the $[0.05-1]\, \rv$ radial range with relative ease \citep{Bartalucci2017}.

Numerical simulations of structure formation in the $\Lambda$CDM cosmology make a number of observationally-testable predictions, such as for instance the existence of a quasi-universal, cuspy, dark matter density profile \citep[e.g.][]{Navarro1997}. However, existing simulations are poorly-adapted to the specific case of high-mass, high redshift systems, as often their resolution is inadequate to match the observations, and/or the number of simulated objects is limited. The present study expands upon the existing body of work on the structure of massive dark matter haloes  \citep[e.g.][]{Tasitsiomi2004,Gao2012,Wu2013} by increasing the spatial or mass resolution and the number of objects. More fundamentally, we simulate the most massive galaxy clusters at $z>0$, and not just the progenitors of the $z=0$ systems. We find that the dark matter profiles of these systems are strikingly similar from $z\sim1$ to the present day, and exhibit a low dispersion (0.15 dex within $\rv$). 

This paper is organised as follows. We briefly introduce the new suite of simulations in Section~\ref{sec:sims}, we discuss the structural evolution from $z=1$ to the present day in Section~\ref{sec:struct} and we discuss our results and conclude in Section~\ref{sec:sum}. Throughout the paper, $M_{\Delta}$ is the mass within radius $r_{\Delta}$, the radius within which the mean mass density is $\Delta$ times the critical density at the cluster redshift.


\section{Simulations and data processing}
\label{sec:sims}

We tailored our simulations to produce a moderately-large sample ($\sim 50$) of massive ($M_{500}\gtrsim5\times10^{14}~\textrm{M}_{\odot}$) objects at $z\sim1$, comparable to current  SZ observational surveys in terms of mass, and with a sample size sufficient to derive robust statistical conclusions. The deepest current observational survey follow-up data sets resolve the inner structure at the tens of kiloparsec scale \citep[e.g.][]{Bartalucci2017}. A simulation reproducing these characteristics would require a $\gtrsim1~\textrm{Gpc}^3$ volume to be simulated at high resolution.
This being impossible given current computational resources, we adopted a strategy in which the total volume was split into three periodic boxes of $1\,h^{-1}$ comoving Gpc on a side containing $2048^3$ dark matter particles, and then used the `zoom-in' technique to further refine individual systems from these parent large box simulations.

The cosmological parameters were taken from \citet{Planck2015} with \{$\Omega_{m}$, $\Omega_{b}$, $\Omega_{\Lambda}$, $\sigma_{8}$, $n_{s}$, $h$\} = \{0.3156, 0.0492, 0.6844, 0.831, 0.9645, 0.6727\}. The initial conditions were generated at a starting redshift of $z=100$ using \music~\citep{Hahn2011} in second-order Lagrangian perturbation theory mode and a transfer function computed with the January 2015 version of the Boltzmann code \textsc{camb}\footnote{http://camb.info/} \citep{Lewis2000}.
The simulations were carried out with the Eulerian adaptive mesh refinement (AMR) code \ramses~\citep{Teyssier2002} with an initial level of refinement $\ell=11~(2048^3)$. Six additional levels of refinement were significantly triggered during the run ($\ell_{max}=17$). This corresponds to a spatial resolution of $\sim15~h^{-1}$ comoving kpc (ckpc; computed as $r_{min}=2\times L_{box}/2^{\ell_{max}}$ i.e.\ corresponding to two times the sidelength of the AMR cells of level $\ell_{max}$). This estimate of the spatial resolution is very conservative. A quasi-Lagrangian refinement strategy was used (i.e.\ the cells were split if they contained more than eight particles). Haloes were identified using \textsc{phew}, the on-the-fly halo finder implemented in \ramses~\citep{Bleuler2015}. Each halo was recentred using a shrinking sphere procedure \citep[e.g.][]{Power2003} and then spherical overdensity masses $M_{\Delta}$ were computed. 

The `zoom' initial conditions were also generated using \music~for a sphere of $8~h^{-1}$ cMpc radius, at the selection redshift $z_{sel}$ defined below, centred on the selected system. However, if the central halo was contaminated by low-resolution particles at more than the $\sim10^{-3}$ level, the simulation was re-run using a  sphere of $16~h^{-1}$ cMpc radius instead. This only affects one system at $z_{sel}=1$ system and one at $z_{sel}=0.8$. We needed an effective resolution of $8192^3$ (8K) particles (see Appendix~\ref{app:res}), resulting in a particle mass of $m_{cdm}=1.59\times10^8~h^{-1}~\textrm{M}_{\odot}$. The main haloes thus contain at least $\sim2$ million particles within $\rv$. We used the same quasi-Lagrangian refinement strategy as for the large boxes, resulting in $\ell_{max}=19$, corresponding to spatial resolutions of $\sim3.8~h^{-1}$ ckpc.

Here we focus on the dark matter only (DMO) `zooms' of the 25 most massive systems selected in four different redshift slices ($z_{\rm sel}=1,~0.8,~0.6$ and 0\footnote{This redshift distribution has been chosen to match an observational sample at $z\gtrsim0.5$ (Arnaud et al. in preparation, \citealt{Bartalucci2017}) as well as providing a `local reference' for comparison with both observations and previous theoretical work.}). The main characteristics of the subsample at each $z_{sel}$, including the resulting minimum mass, are summarised in Table~\ref{tab:25most}. The sample grows by nearly a factor of four in median mass between $z=1$ and $0$. The relaxed fraction $f_{\rm rel}$ is computed as the fraction of systems with $\Delta r\leq0.04$ \citep{Power2012}, where $\Delta r$ is the distance between the centre of mass within the \citet{Bryan1998} virial radius $r_{vir}$ and the centre of the shrinking sphere, in units of $r_{vir}$.  A resolution study shows that the 8K zooms (i.e.\ the nominal resolution) are converged (in particular with the even-higher resolution 16K runs) over the whole resolved ($r\ge r_{\rm min}^{\rm 8K}$) radial range (see Appendix~\ref{app:res}) and that an effective resolution of $8192^3$ is the minimum required for this work. 

\begin{table}
\centering
\caption{Main characteristics of the 25 most massive sample for each redshift slice.}
\label{tab:25most}
\begin{tabular}{p{1.3cm}p{1.3cm}p{1.55cm}p{1.3cm}c} 
\hline
Redshift slice & $\min(M_{500})$ [$10^{14}~\textrm{M}_{\odot}$] & median$(M_{500})$ [$10^{14}~\textrm{M}_{\odot}$] & $\max(M_{500})$ [$10^{14}~\textrm{M}_{\odot}$] & $f_{rel}$\\
\hline
$z_{sel}=1$ & $5.55$ & $6.32$ & $13.9$ & $24\%$ \\
$z_{sel}=0.8$ & $7.27$ & $8.31$ & $12.8$ & $16\%$ \\
$z_{sel}=0.6$ & $8.97$ & $10.6$ & $30.0$ & $32\%$ \\
$z_{sel}=0$ & $19.2$ & $22.6$ & $36.5$ & $40\%$ \\
\hline
\end{tabular}
\end{table}

\begin{figure}
\includegraphics[width=1.0\columnwidth]{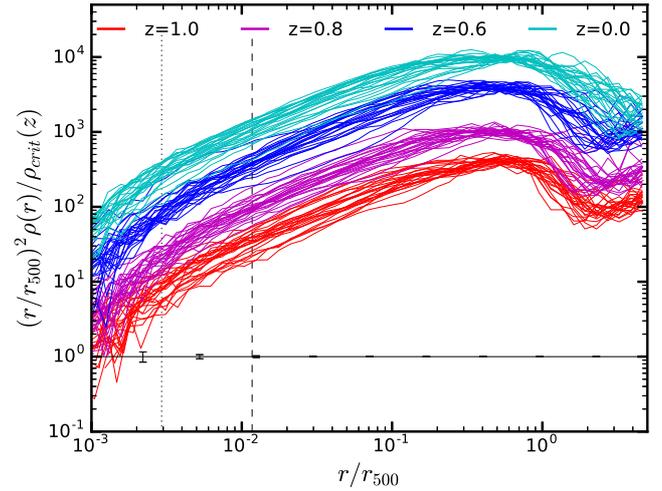}
\caption{Spherical density profiles for each redshift slice. For clarity, the profiles of the $z_{sel}=1,~0.8,~0.6$ and 0 systems have been shifted up by a factor of 2, 5, 20 and 50, respectively. The resolution limit at $z=0$ is depicted as dashed and dotted vertical lines for the large boxes and the DMO zooms, respectively. The typical Poisson errors for every fifth data point are displayed as black error bars.}
\label{fig:rhoprof}
\end{figure}

\begin{figure*}
\includegraphics[width=1.0\columnwidth]{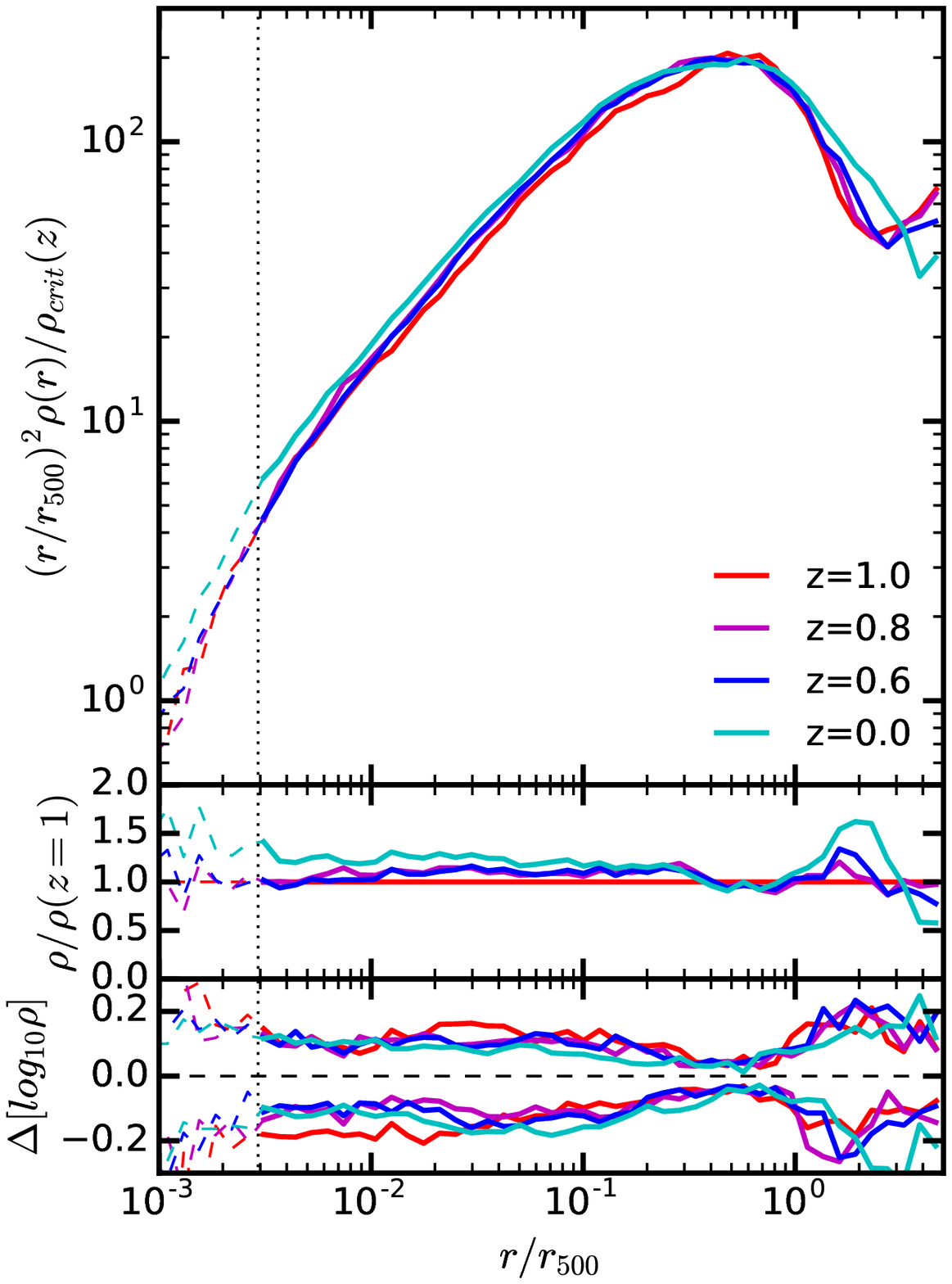}
\includegraphics[width=1.0\columnwidth]{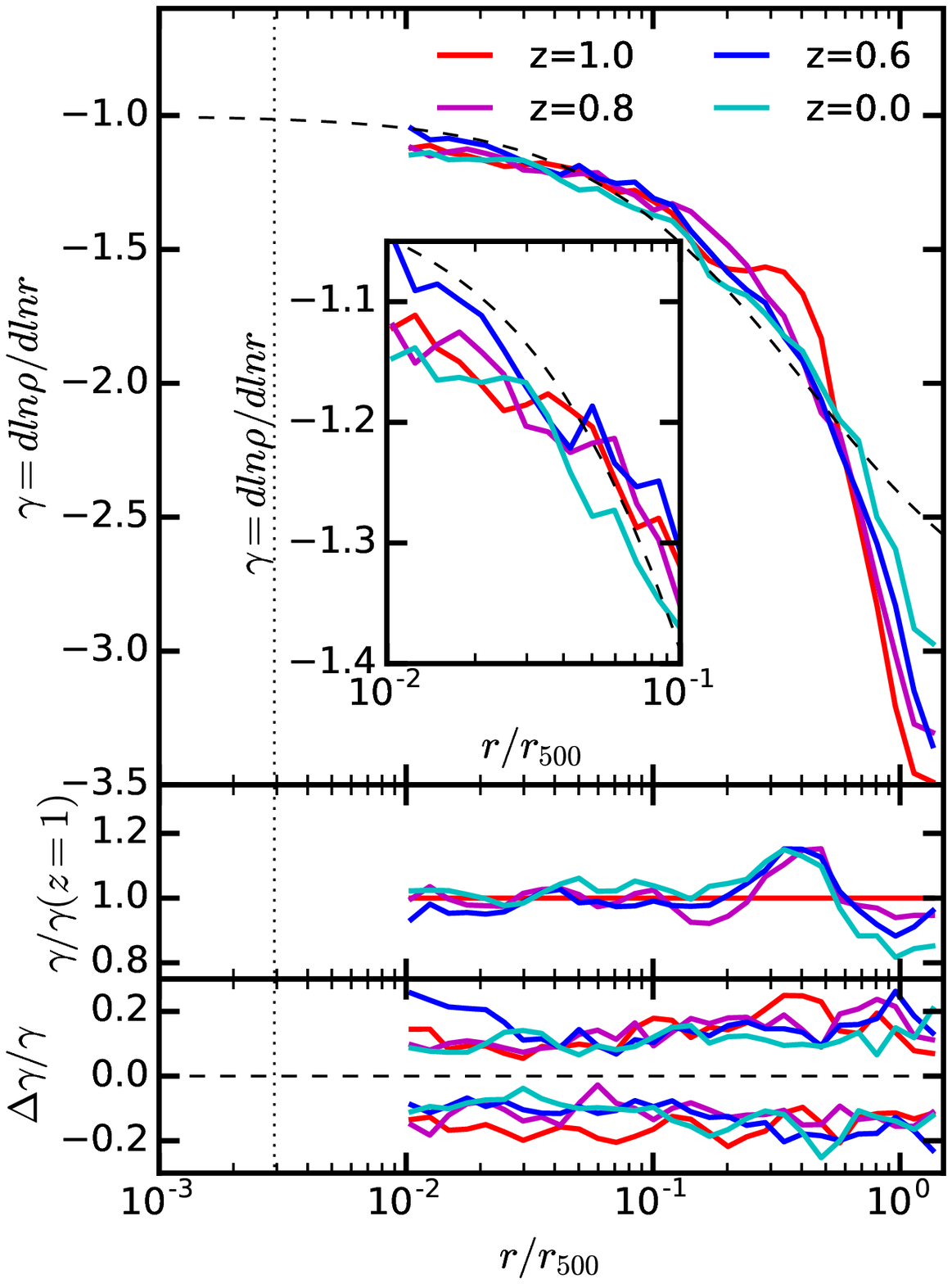}
\caption{Median spherical density (\emph{left}) and logarithmic slope (\emph{right}) profiles for each redshift slice (\emph{top}). The \emph{middle} panels show the median profiles normalised by the median profile at $z_{sel}=1$ whereas the \emph{bottom} panel displays the difference between the 16/84 percentiles and the median. In all panels, the resolution limit at $z=0$ is depicted as a dotted line and the profiles are plotted either as dashed thinner lines below that limit for the density or only from the eighth radial bin above that limit for the logarithmic slope where the Savitzky-Golay filter method is valid. The inset in the \emph{top right} panel is a zoom of the $0.01\le r/\rv\le0.1$ radial range.}
\label{fig:rhoprofbis}
\end{figure*}

Figure~\ref{fig:rhoprof} shows the density profiles computed using 50 evenly-spaced logarithmic bins over the $10^{-3}\leq r/\rv\leq5$ radial range with the \textsc{pymses}\footnote{http://irfu.cea.fr/Projets/PYMSES/intro.html} python module.  We use mass-weighted radii in  this and subsequent figures. The resolution limits at $z=0$ (the maximum value of $r_{\rm min}/r_{500}$) are displayed as black dashed and dotted lines for the large boxes and the DMO zooms, respectively. Both limits are nearly redshift-independent as, with increasing redshift, the decrease in the minimum value of $r_{500}$ is nearly compensated by the increase of $r_{min}$ given in comoving units. The `zoom-in' procedure results in the gain of at least a factor of five in spatial resolution.  Being smaller than the Poisson errors, the fluctuations are real and are likely due to the presence of substructures, and to oscillations resulting from the relaxation of the haloes. 

Densities were normalised by the critical density of the Universe at the cluster redshift, and  the radii by the corresponding $\rv$. This choice was motivated by the fact that the inner profiles are expected to evolve most self-similarly when rescaled in terms of the critical density \citep{Diemer2014}, and that the  region $r/\rv\lesssim1$ can be routinely probed by observations.
We multiplied the scaled  densities by $(r/\rv)^2$ to reduce the dynamic range of the $y$-axis. To  quantify the profile shapes, we computed the logarithmic slope $\gamma\equiv \textrm{d}\ln\rho/\textrm{d}\ln r$ from the smallest resolved radius ($\max(r_{min}/\rv)\lesssim0.003)$,  using the fourth-order Savitzky-Golay algorithm over the 15 nearest bins \mbox{\citep{Savitzky1964}}.  Excluding the seven innermost and outermost bins where the method is no longer valid, the logarithmic slope profiles cover  the $0.01\leq r/\rv \leq 1.35$ radial range.  
 

\section{Structural evolution from $z=1$}
\label{sec:struct}

The left panels of Fig.~\ref{fig:rhoprofbis} show the median scaled density profiles of each redshift slice. In the standard  self-similar model, we expect the density profiles to be identical, irrespective of the cluster redshift or any other characteristics. Hereafter, evolution refers to the evolution of the scaled profiles, i.e.  additional evolution in excess of the  self-similar expectation. The mean density within  $\rv$ being  proportional to  $\rho_{\rm crit}(z)$ by definition,  scaled profiles  may only differ by their shape  in the region $r/\rv<1$.  Additionally, they are expected to cross around  $r/\rv\sim0.6$, the radius enclosing half of $M_{500}$,  and thus their scatter to be minimal around that radius.  Beyond $\rv$,  a breaking of self-similarity could translate  into evolution in both shape and overall  normalisation.

The scaled profiles display a barely noticeable evolution for $1\geq z\geq 0.6$ over the whole radial range probed by the simulations. They change  by never more than a factor of $\sim1.2$ with decreasing redshift (see the \emph{middle} panel).  Conversely, they evolve a bit more (by up to a factor of $\sim1.5$) both in the core ($r/\rv\lesssim0.2$) and the outskirts ($r/\rv\gtrsim1.5$) for $z\leq0.6$. This is consistent with the `stable clustering' hypothesis often used for computing the non-linear matter power spectrum \citep[e.g.][]{Peebles1980}. The former evolution corresponds to a slight  increase of  peakiness. The latter is most likely mostly due to their mass growth. The evolution in the outskirts corresponds to the transition between the one- and two-halo terms moving outwards as the haloes grow. It is thus especially noticeable as the median masses of the respective samples more than double from $M_{500}=1.06\times10^{15}~\textrm{M}_{\odot}$ to $M_{500}=2.26\times10^{15}~\textrm{M}_{\odot}$ between $z=0.6$ and $z=0$. 
Note that up to between 42 (between $z=1$ and 0) and 83 per cent (between $z=1$ and 0.8) of this mass growth could be due to pseudo-evolution \citep[e.g.][]{Diemer2013,Wu2013}.  

The scatter of the density profiles, as depicted in the \emph{bottom} panel,  is remarkably small over the  redshift and radial ranges ($\lesssim0.2$ dex).  The scatter is less than $0.15$ dex within $\rv$. In  the outskirts, it is slightly larger  than in the core  and seems to increase mildly (by $\sim30$ per cent) with redshift. Note that these trends need to be confirmed with the whole set of simulations as each redshift slice only contains 25 systems. The minimum  around  $r/\rv\sim0.6$ is a consequence of the normalisation of the profiles as mentioned above.

\begin{figure}
\includegraphics[width=1.0\columnwidth]{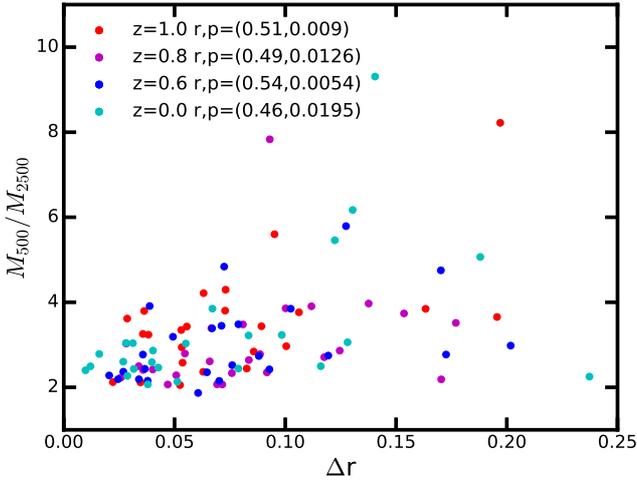}
\caption{$M_{500}/M_{2500}$ as a function of relaxation state for each redshift slice. The Spearman's rank and the null hypothesis probability are listed in the legend.}
\label{fig:sparsity}
\end{figure}

The high degree of self-similarity is confirmed when studying the profile shapes. The right panels of Fig.~\ref{fig:rhoprofbis} show the median logarithmic slope profiles of each redshift slice. They exhibit small amounts of both evolution (it never exceeds 20 per cent; see the \emph{middle} panel) and dispersion ($\lesssim0.2$ dex; see the \emph{bottom} panel) over the whole redshift and radial range. Similarly to the density profiles, the evolution is more important in the very central  regions ($r/\rv\lesssim0.02$) and the outer regions ($r/\rv\gtrsim0.6$).  More unexpectedly, the slope of the profiles, contrarily to the profiles themselves, displays noticeable evolution for all $z\leq1$ and not just  $z\leq0.6$. \\

The scatter of the density slope displays nearly no evolution over the whole radial range and is similar in amplitude to that of the density profiles (except around $r/\rv\sim0.6$, where the scatter of the density is minimum by construction).  More importantly, the inner slope shows no signs of converging to an asymptotic value but is still consistent with a NFW profile of typical concentration for massive galaxy clusters ($c_{200}=3.5$) over the radial range in question (see the \emph{top} panel and its inset; in agreement with the results of e.g.\ \citealt{Navarro2004}). 

A first investigation of the origin of the scatter in the density profiles is illustrated by Fig.~\ref{fig:sparsity}, which displays the $M_{500}/M_{2500}$ ratio as a function of $\Delta r$. The most relaxed clusters ($\Delta r\leq0.04$) tend to be more centrally concentrated, i.e. they have a smaller $M_{500}/M_{2500}$ ratio ($M_{500}/M_{2500}\lesssim4$). However,  there is no clear correlation between both parameters   in terms of Spearman's rank and null hypothesis probability,  as listed in the legend of Fig.~\ref{fig:sparsity}. In fact, the distribution is consistent with all systems having a similar $M_{500}/M_{2500}$ ratio, irrespective of their relaxation state, but with a dispersion that increases with increasing $\Delta r$. Therefore, the scatter in the density profiles is only connected to the relaxation state of the galaxy cluster  through the fact that relaxed clusters are mostly centrally concentrated, while  unrelaxed objects span a larger variety of profile shapes, including very shallow profiles. Note that the entire simulation sample is required to reach a reliable conclusion on this point.


\section{Discussion and conclusion}
\label{sec:sum}

We have focussed in this work on the dark matter density profiles of the 25 most massive systems selected from DMO simulations in four redshift slices ($z_{\rm sel}=1,~0.8,~0.6$ and 0).  With a median mass of $M_{500} = 6.3\times10^{14} M_{\odot}$ at $z=1$, the sample is composed of the rarest objects, probing for the first time the extreme limits of the cluster mass function, such as would be detectable observationally only in all-sky surveys. Surprisingly, these objects exhibit a high level of self-similarity, and their dark matter density profiles can be described with the typical NFW profile found in relaxed local systems. 

From the Millenium simulations \citep*{Fakhouri2010},  high--mass systems with  $M>5\times 10^{14}~\textrm{M}_{\odot}$  at $z=1$ ($z=0$)  have undergone at least one major merger (mass ratio $>1:3$) during the preceding 4 Gyr (12 Gyr). The relaxation time estimated from their crossing time ($t_{cross}\propto \rv/\sigma_{500}$ with $\sigma_{500}=(GM_{500}/\rv)^{1/2}$) is close to two Hubble times ($t_{H}=1/H(z)$), i.e.\ about 16 Gyr (29 Gyr). A similar conclusion is reached  if one uses the dynamical time.  
We expect that these objects should still be forming and thus be highly unrelaxed.
Using $\Delta$r  as a dynamical indicator, or from visual inspection of the images, we find that the vast majority of the systems in question are indeed unrelaxed.

Naively, one would then expect that the  density profiles of such objects would exhibit large variations, linked to the wide variety of dynamical states and formation histories, and that the median profile has not yet converged to the near-universal form of relaxed systems in the local Universe. In contrast,  once scaled according to the critical density at each redshift, the density profiles of the clusters in our simulations are remarkably similar, with a low dispersion of less than 0.15 dex within $\rv$. Furthermore, there is little evolution in  the radial logarithmic slope or scatter with redshift. This surprising result suggests that the `universal', `broken/running' power-law, density profile (e.g.\ similar to NFW or Einasto)  is already in place at $z>1$ and that it is robust to merging activity. This conclusion is similar to that recently  obtained   for primordial (Earth--mass)  haloes  by \citet{Angulo2016} and \citet{Ogiya2016}, but at scales that are 21 orders of magnitude larger. 

Interestingly, \citet{McDonald2017}  recently found  a remarkably standard self-similar evolution  in the mean profile of the hot gas beyond  the cooling core region in massive clusters up to $z\sim1.9$.  This would be  a natural consequence of the self-similar evolution of the underlying dark matter distribution that we have shown here, since  these systems are dark-matter dominated and the gas evolution, except in the very central regions,  is dominated by simple gravitational  physics.  One could even speculate that McDonald et al.'s finding that the cool core formed early ($z\gtrsim1.5$) is made possible by the early establishment and stability over time of the centrally--cusped dark matter profile. However, the actual demographics of the core properties and their evolution (e.g.\ the nearly constant cool core mass found by McDonald et al.) will also depend on the specific gas physics (e.g.\ shocks, cooling, AGN feedback).

There is an indication of a residual link between the profile shape and dynamical state, with the most unrelaxed clusters exhibiting a larger dispersion.   Future work on a larger sample, covering a wider mass range, will enable us to better characterise both the evolution and the scatter. These are essential to understand the link between profile shape, dynamical state, and formation history.

\section*{Acknowledgements}
The authors thank the anonymous referee for a constructive report. This work was supported by the French Agence Nationale de la Recherche under grant ANR-11-BS56-015 and by the European Research Council under the European Union's Seventh Framework Programme (FP7/2007-2013) / ERC grant agreement number 340519. It was granted access to the HPC resources of CINES under the allocations 2015-047350, 2016-047350 and 2017-047350 made by GENCI. AMCLB is extremely grateful to Damien Chapon for his help with \textsc{pymses} and Patrick Hennebelle for granting access to his shared memory machine. She further thanks for its hospitality the Institute for Computational Science of the University of Z\"urich. 




\bibliographystyle{mnras}
\bibliography{25mostmassive} 



\appendix

\section{Resolution study}
\label{app:res}

\begin{figure}
\includegraphics[width=1.0\columnwidth]{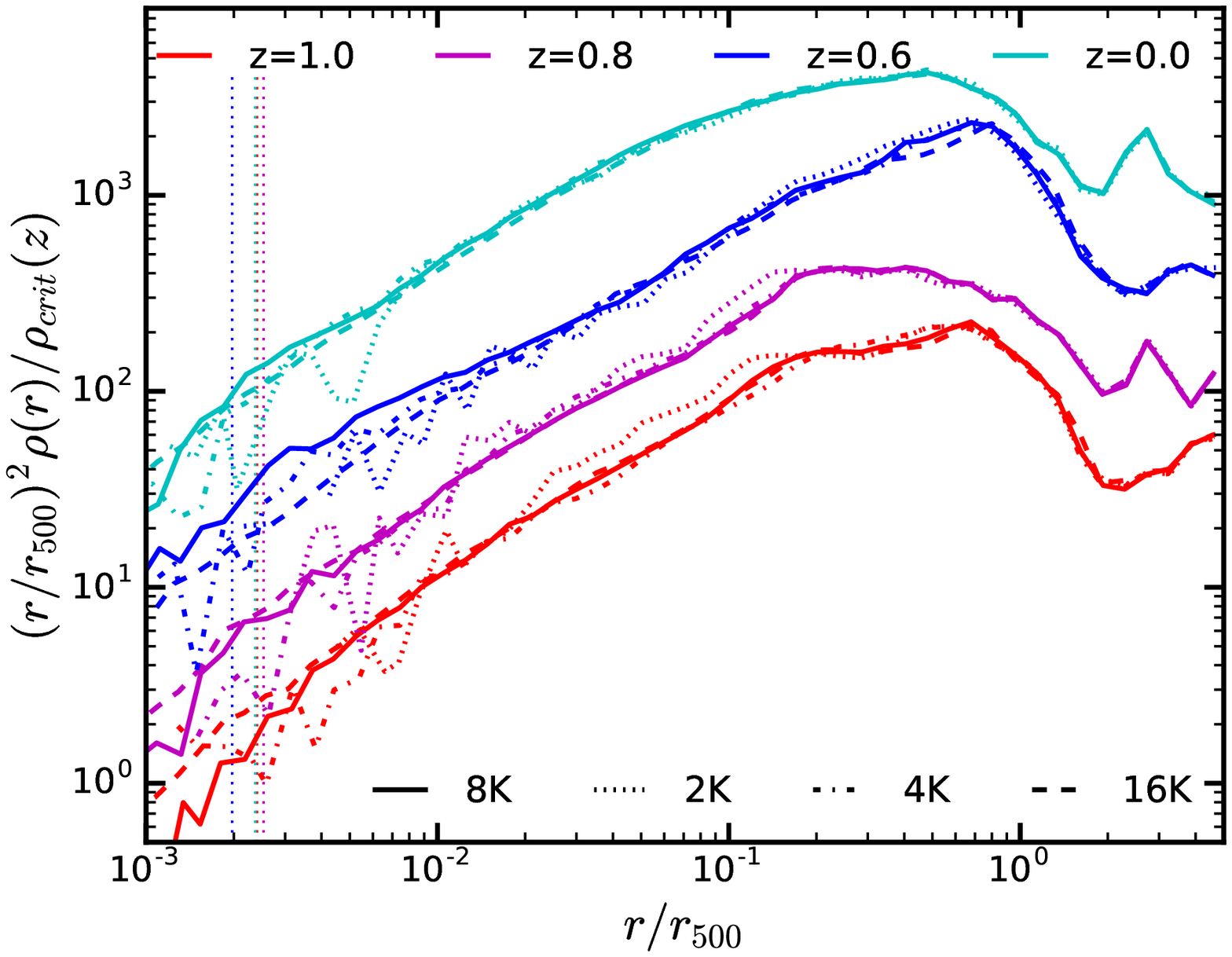}
\caption{Resolution test for the spherical density profiles for the most massive system of each redshift slice. For clarity, the profiles of the $z_{\rm sel}=0.8,~0.6$ and 0 systems have been shifted up by a factor of 2, 10 and 20, respectively. The 2K, 4K, 8K (nominal resolution) and 16K are shown as dotted, dash-dotted, solid and dashed lines, respectively. The resolution limits $r_{\rm min}$ of the 8K runs are drawn as vertical coloured dotted lines.}
\label{fig:restest_rho}
\end{figure}

Fig.~\ref{fig:restest_rho} shows the results of a resolution study for the most massive system of each redshift slice. The 8K (i.e.\ the production runs) and 16K density profiles are converged over the whole resolved ($r\geq r_{\rm min}^{\rm 8K}$) radial range for all the redshift slices. This conclusion applies for all the systems for which resolution tests have been run so far (and not just the most massive system of each redshift slice). As the 2K and 4K profiles are significantly different from the 8K and 16K over the interesting radial range (i.e.\ below $10-15$ kpc and down to a few kpc that is equivalent to $r/\rv\lesssim1.5\times10^{-2}$), an effective resolution of $8192^3$ is the minimum required for this work.


\bsp	
\label{lastpage}
\end{document}